\documentclass[12pt]{article}
\usepackage{amssymb,amsmath,amsthm,dsfont}

\theoremstyle{plain}
\newtheorem{thm}{THEOREM}

\theoremstyle{definition}

\newcommand{\Hh}{{\mathcal H}}
\newcommand\Ee{{\mathcal E}}
\newcommand{\F}{{\mathcal F}}

\newcommand{\Pp}{{\mathcal P}}
\newcommand{\R}{{\mathbb R}}

\newcommand\infspec{{\rm inf\, spec\,}}

\begin{document}

\title{Equivalence of two definitions of the effective mass of a
  polaron} \author{\vspace{5pt} Elliott H.~Lieb$^{*}$ and Robert
  Seiringer$^{
    \dagger}$\\
  \vspace{-4pt}\small{$^{*}$Departments of Mathematics and Physics,
    Jadwin Hall, Princeton University} \\
  \small{Washington Road, Princeton, New Jersey 08544-0001, USA}\\
  \vspace{-4pt}\small{$^{\dagger}$Department of Mathematics and
    Statistics, McGill University}\\ \small{805 Sherbrooke St. West,
    Montreal QC H3A0B9, Canada}} 

\date{\small April 5, 2013}

\maketitle

\textit{Dedicated to Herbert Spohn, a leader in the mathematical study of
the polaron, on the occasion of his retirement from T.U. M\"unchen}
\bigskip

\renewcommand{\thefootnote}{$ $}
\footnotetext{\copyright\, 2013 by the authors. This paper may be reproduced, in its
entirety, for non-commercial purposes.}

\begin{abstract}
  Two definitions of the effective mass of a particle interacting with
  a quantum field, such as a polaron, are considered and shown to be
  equal in models similar to the Fr\"ohlich polaron model. These are:
  1. the mass defined by the low momentum energy $E(P) \approx E(0)+
  P^2/2M $ of the translation invariant system constrained to have
  momentum $P$ and 2. the mass $M$ of a simple particle in an
  arbitrary slowly varying external potential, $V$, described by the
  nonrelativistic Schr\"odinger equation, whose ground state energy
  equals that of the combined particle/field system in a bound state in the same
  $V$.
\end{abstract}

\section{Introduction}

When a particle, such as an electron, clothes itself with a quantized
field of some kind, it is supposed to behave, to a good approximation,
like a particle described by a different mass and, perhaps, a
different charge.  This belief notwithstanding, it has yet to be seen
clearly, explicitly and non-perturbatively, how one can deduce the
experimental Balmer lines of hydrogen from a fully interacting theory
of quantum electrodynamics without cut-offs.

Herbert Spohn studied this question from a dynamical point of view \cite{Sb,TS},
showing that the dynamics of a dressed particle can be approximated to
a certain extent by a simpler effective dynamics (see also \cite{BC,T,TT}). This work is typical
of his profound analysis of many fundamental aspects of theoretical
physics. We study the same problem from a different, stationary,
aspect, which is in some ways simpler and therefore might be of
interest.  We consider here the polaron model of H. Fr\"ohlich
\cite{Fr}, which describes an electron of bare mass $m$ interacting
with the quantized electric field of dipoles in a polar crystal; in
contrast to quantum electrodynamics, there are no infinities in this
theory that require renormalization.  Spohn has pioneered much of the
theoretical understanding of the polaron, especially its effective
mass \cite{Fe, Pe, Sp}, and we hope these remarks may encourage him to
continue his engagement with the subject.

The coupling constant in the model is denoted by $\alpha$. When
$\alpha$ is small perturbation theory does seem to be adequate, but
usually $\alpha$ is rather large.  For very large $\alpha$, the
approximate non-linear theory of Pekar \cite{Pe} applies, at least to
describe the ground state energy \cite{DV, LT}. The diameter of the
polarization field cloud around the electron is proportional to
$\alpha^{-1}$ in Pekar's approximation.

We now place this object in a potential $V$ that varies little over
dimensions corresponding to the size of the cloud, and we ask the
following question: Is it true that the ground state energy equals the
polaron ground state energy, $E(0)$, plus the non-relativistic
Schr\"odinger ground state energy of a particle of mass $M$ in the
potential?  Here, $M$ is the effective mass of the polaron defined for
the translation invariant problem ($V=0$) by the bottom of the
spectrum in the momentum $P$ fiber, $E(P)\approx E(0) + P^2/2M$, as
$P\to 0$. In other words, can we define the effective, renormalized
mass of a particle in two different, but equivalent ways: in terms of
the ground state energy of the translation invariant problem (in the
fiber of fixed total momentum $P$) or in terms of the ground state
energy of the particle in a slowly varying potential well, such as an
electron in the ground state of a hydrogen atom (as was attempted for
nonrelativistic QED in \cite{LL})?

In this paper we show how this can be proved for the polaron, for any
$\alpha$. Our method applies to a general class of models, as will be
discussed after Theorem~\ref{t1} in the next section.  Many open
questions remain and some of these are reviewed at the end of the
paper in Section~\ref{sec:end}.

\section{Definition of the Problem and Main Result}

The Hilbert space for a single polaron is 
$$
\Hh = L^2(\R^d) \otimes \F 
$$
where $\F$ is the bosonic Fock space over $L^2(\R^d)$. Physically, it
represents the electric field produced by the creation of dipoles in the
optical
mode of a polar crystal.

The Hamiltonian of the polaron is
\begin{equation}\label{ham}
H = p_x^2 + H_f + \int_{\R^d} v(k) e^{-ikx} a^\dagger_k dk + \int_{\R^d}
\overline{v(k)} e^{ikx} a_k dk,
\end{equation}
with $p_x= -i\nabla_x$ (in units in which $\hbar=1$ and the particle mass equals $m=1/2$), and with
field energy
$$
H_f = \int_{\R^d} \omega(k) a^\dagger_k a_k dk.
$$
The momentum of a mode is denoted by $k\in \R^d$. For Fr\"ohlich's
polaron, $d=3$ and the energy of a mode, $ \omega(k)$, is a positive
constant, independent of $k$.   Moreover, the interaction $v(k)\propto \sqrt{\alpha}
|k|^{-1}$ in this model. 
For our purposes here, however, we can consider general dimensions $d$ as well as more general functions $\omega(k)$ and $v(k)$. We only require, at this point, that $H$ defines a self-adjoint operator that is semi-bounded from below.

The important thing is that $H$ commutes with total momentum
$\mathcal{P}=p_x+ P_f$, where $P_f = \int_{\R^d} k a^\dagger_ka_k dk$.
There is the well known fiber decomposition in which $H$ is restricted to 
states of a fixed numerical value, $P$, of $\mathcal{P}$, namely
\begin{equation} \label{fibre}
H_P = (P - P_f)^2 + H_f + \int_{\R^d} v(k) a^\dagger_k dk + \int_{\R^d}
\overline{v(k)}  a_k dk,
\end{equation}
which acts on $\F$ alone. Then
$$
H \cong \int_{\R^d}^\oplus H_P\, dP\,.
$$

Define
$$
E(P) = \infspec H_P\, ,
$$
and define the \lq\lq dynamic effective mass\rq\rq $M$, which is greater than $m=1/2$, by
\begin{equation}\label{dyn:mass}
\frac 1 M = 2 \lim_{P\to 0} \frac{E(P)-E(0)}{|P|^2}\,.
\end{equation}
We assume that this limit exists, but $M$ need not be finite, a priori.   
For Fr\"ohlich's polaron it is known that $1/2< M<\infty$ for all $\alpha>0$
\cite{Ju,GL}, i.e., there is no self-trapping of the electron.

To define the \lq\lq static effective mass\rq\rq $\widetilde M$ we let
$V$ be a bounded, real-valued function that decays at infinity and has
the property that $p_x^2 + V(x)$ has a negative energy bound
state. (For technical reasons, we assume slightly more regularity,
namely that the Fourier transform $\hat V$ is in $L^1(\R^d)$.)  Then
the function
$$
\Ee:[1/2,\infty) \to (-\infty,0) \quad , \quad \Ee(m) = \infspec \left(
\frac{p_x^2}{2 m} + V(x)\right)
$$
is negative and strictly decreasing for $m\geq 1/2$, and hence is invertible. 

Note that, by scaling, $p_x^2/2m + \lambda^2 V(\lambda x)$ for $\lambda>0$ is unitarily equivalent to $\lambda^2(p_x^2/2m + V(x))$. 
The definition of $\widetilde M$ is
$$
\Ee(\widetilde M) = \lim_{\lambda \to 0} \frac 1 {\lambda^2} \infspec \left(
H + \lambda^2 V(\lambda x) - E(0) \right) \,.
$$
The purpose of $\lambda $ here is merely to stretch the scale of $V$
in order to insure that
variations in $V(x)$ have much longer wavelength than relevant modes of the
field.

Our main result is the following.
\begin{thm} \label{t1}
 Under suitable conditions on
$v$ and $\omega$ (which are satisfied for Fr\"ohlich's polaron model)
\begin{equation}\label{MR}
\boxed{ \phantom{_x} M = \widetilde M\!\!\phantom{_{x_y}} }
\end{equation}
\end{thm}

To be precise, our theorem applies to all models of the form (\ref{ham}), satisfying the properties
\begin{enumerate}
\item [(1)] The limit (\ref{dyn:mass}) exists, and there is a constant $C>0$ such that $E(P) \geq E(0) + P^2[ 2 M (1+ C P^2)]^{-1}$
\item [(2)] There exists a $P_c>0$ such that for all $P$ with $|P|<P_c$, $H_P$ in (\ref{fibre}) has a unique ground state, which depends continuously on $P$
\end{enumerate}
These Assumptions are known to be fulfilled for the Fr\"ohlich
polaron, see \cite{GL} or \cite{Mo}. Assumption (1) implies, in
particular, that $E(P) \geq E(0)$ for all $P\in \R^d$, and takes
account of the fact that, in general, the function $E(P)$ does not
increase unboundedly. It is bounded above by $\omega(P)$, in
fact. This is seen from \eqref{fibre} where we can obtain a
variational upper bound to $E(P)$ by using the state consisting of
just one phonon of momentum $k=P$.  Our proof of Theorem~\ref{t1} is
slightly complicated by the fact that we need to take account of the
boundedness of $E(P)$.

The proof of Theorem~\ref{t1} consists
of two steps. In Section~\ref{sec:geq}, we shall show that Assumption
(1) above implies that $M\geq \widetilde M$. The reverse inequality
$M\leq \widetilde M$ follows from Assumption (2) (and the first part
of Assumption (1)), as shown in Section~\ref{sec:leq}.

We remark that our method can be generalized to settings where
Assumption (2) does not necessarily hold. For instance, for models of
non-relativistic QED, $\Phi_P$ does not exist for $P\neq 0$ unless
one introduces an infrared cut-off \cite{Ju} (in addition to the
necessary ultraviolet cut-off). One can either apply our method
directly to a model with infrared cut-off, and then argue that the
effective mass is continuous in the cut-off \cite{pf}. Alternatively,
one could work with the full model (without infrared cutoff) but take
as functions $\Phi_P$ in our variation argument in
Section~\ref{sec:leq} the ground states of the model with a suitable
$P$-dependent cut-off. Given enough control of the dependence of these
functions on $P$ and the cut-off (as examined in \cite{piz} for the
Nelson model for weak coupling) our argument applies.

\section{$M\geq \widetilde M$}\label{sec:geq}

With $\mathcal{P}=p_x+ P_f$ denoting the total momentum operator on $\Hh$,
we have 
$$
H \geq E(\Pp).
$$
This is so because, for each fiber, the number  $E(P) $ is the bottom of the 
spectrum. This function $E$ is used to define
the {\it operator} $ E(\Pp)$, which is unitarily
equivalent to the operator $E(p_x)$, via the unitary
transformation $U=e^{  ix P_f}$. We note that $U^* x U =x$. 
Hence 
$$
 \infspec \left( H + \lambda^2 V(\lambda x) - E(0) \right) \geq  \infspec 
\left( E(p_x) + \lambda^2 V(\lambda x) - E(0) \right)
$$
and, therefore,
$$
\Ee(\widetilde M) \geq \lim_{\lambda\to 0} \infspec\left( \frac{ E(\lambda p_x) - E(0)}{\lambda^2} + V(x) \right), 
$$
where we applied a unitary rescaling of $x$ by $\lambda$.

Let us assume that 
\begin{equation}\label{assu}
E(P) \geq E(0) + \frac{ P^2}{2M(1+ C P^2)}
\end{equation}
for some $C>0$. 
We claim that 
\begin{equation} \label{desire}
\infspec\left( \frac{ E(\lambda p_x) - E(0)}{\lambda^2} + V(x) \right) \geq
\Ee(M) - O(\lambda)
\end{equation}
as $\lambda\to 0$, which implies that $\Ee(\widetilde M) \geq \Ee(M)$, hence
$M\geq \widetilde M$ since $\Ee$ is a decreasing function.

For $\beta>0$, let $\chi(P) = \theta(\beta - |P|)= 1$ if $|P| \leq
\beta$ and $=0$ otherwise,  and  let $\tilde\chi(P) = 1 - \chi(P)$. From (\ref{assu}) we
have
$$
E(P) - E(0) \geq \chi(P)  \frac{P^2}{2M(1+ C \beta^2)} + \tilde
\chi(P) \frac{\beta^2}{2M(1+C\beta^2)}\,.
$$
We split $V = V_+ - V_-$ into its positive and negative parts (i.e.,
$V_+(x) = \max\{V(x),0\}$),  and use Schwarz's inequality to bound
$ \chi(\lambda p_x) V_\pm(x) \tilde\chi(\lambda p_x)$. In this way we conclude that 
\begin{align*}
V(x) & \geq (1-\epsilon) \chi(\lambda p_x) V_+(x) \chi(\lambda p_x)  
+ \left( 1 - \epsilon^{-1} \right) \tilde \chi(\lambda p_x) V_+(x)  
\tilde \chi(\lambda p_x) \\ & \quad - (1+\epsilon) \chi(\lambda p_x) V_-(x) \chi(\lambda p_x) 
 - \left( 1 + \epsilon^{-1} \right) \tilde \chi(\lambda p_x) V_-(x)  \tilde \chi(\lambda p_x)
\end{align*}
for any $\epsilon>0$. We thus obtain
\begin{align*}
& \frac{ E(\lambda p_x) - E(0)}{\lambda^2} + V(x)  \\ & \geq \chi(\lambda p_x)
\left( \frac{p_x^2}{2M(1+C\beta^2)} + (1-\epsilon)V_+(x) - (1+\epsilon)
V_-(x) \right) \chi(\lambda p_x)  \\ & \quad + \tilde\chi(\lambda p_x)
\left( \frac{\beta^2}{2\lambda^2 M(1+C \beta^2)} - (1+\epsilon^{-1})
\|V\|_\infty\right)   \,.
\end{align*}
For the choice $\epsilon = O(\lambda)$ and $\beta = O(\lambda^{1/2})$, one easily sees that 
$$
 \frac{p_x^2}{2M(1+C\beta^2)} + (1-\epsilon)V_+(x) - (1+\epsilon)
V_-(x) \geq \mathcal{E}(M) - O(\lambda)\,.
$$
Moreover, we can choose $\epsilon = C \lambda$ with $C$ large enough such that 
$$
\frac{\beta^2}{2\lambda^2 M(1+C \beta^2)} - (1+\epsilon^{-1})
\|V\|_\infty \geq 0 
$$
for small $\lambda$. 
This gives the
desired result \eqref{desire}.

\section{$M\leq \widetilde M$}\label{sec:leq}

Let $\Phi_P\in \F$ denote the ground state of $H_P$, which we assume
to exist and to be unique up to a phase for $|P|<P_c$. This is
known to be the case for the Fr\"ohlich polaron, see \cite[Statement
1]{GL}.  For $x\in \R^d$, let $\Phi_P^x = \exp\{-ixP_f\} \Phi_P$, and
define $\Psi\in \Hh$ as
\begin{equation}\label{int:psi}
\Psi = (2\pi \lambda)^{-d/2} \int_{\R^d} \hat f(\lambda^{-1}P)\, \exp\{iPx\}\,
\Phi_P^x\,dP,
\end{equation}
in which $f\in L^2(\R^d)$ with $\|f\|=1$ and $\hat f$ is its Fourier
transform (normalized so that $\Vert \hat f \Vert =1$). We assume that
$\hat f$ has compact support, so that for small enough $\lambda$ the
integral in (\ref{int:psi}) is only over $P$'s with $|P|<P_c$, where
we know $\Phi_P$ to exits.  It is readily checked that
$\|\Psi\|=1$. Moreover,
$$
\langle \Psi | H \Psi\rangle = \int_{\R^d} |\hat f(P)|^2 E(\lambda P)\, dP
$$
and
$$
\langle \Psi|  V(\lambda x) \Psi\rangle = (2\pi)^{-d/2}\int_{\R^d\times \R^d} \overline{\hat f(P)} f(P') \hat V(P-P') \langle \Phi_{\lambda P}|\Phi_{\lambda P'} \rangle\, dP\,dP' \,,
$$
where the latter inner product in the integrand is in $\F$. For fixed $f\in H^1(\R^d)$, 
$$
\lim_{\lambda \to 0} \int_{\R^d} |\hat f(P)|^2 \frac{E(\lambda P)-E(0)}{\lambda^2} dP =  \int_{\R^d} |\hat f(P)|^2 \frac{P^2}{2M} dP
$$
by dominated convergence (using the fact that $E(P)\leq E(0)+P^2$). If we
now assume that $\hat V \in L^1(\R^d)$,  we also have that
\begin{align*}
&\lim_{\lambda\to 0} \int_{\R^d\times \R^d} \overline{\hat f(P)} f(P') \hat V(P-P') 
\langle \Phi_{\lambda P}|\Phi_{\lambda P'} \rangle\, dP\,dP' \\ & =  \int_{\R^d\times \R^d} 
\overline{\hat f(P)} f(P') \hat V(P-P') \, dP\,dP'\,.
\end{align*}
Again we used  dominated convergence, together with the continuity of $\Phi_P$, which implies that
$\langle \Phi_{\lambda P}|\Phi_{\lambda P'} \rangle $ converges to 
$\langle  \Phi_{0}|\Phi_{0} \rangle =1 $ as  $\lambda \to 0$. 

We  have, therefore,
\begin{align}\nonumber
\Ee(\widetilde M) & \leq \lim_{\lambda \to 0} \left( \lambda^{-2} \langle \Psi| H - E(0) | \Psi \rangle + \langle \Psi | V(\lambda x) \Psi\rangle \right) \\ &=  \int_{\R^d} |\hat f(P)|^2 \frac{P^2}{2M}\, dP + \int_{\R^d} |f(x)|^2 V(x) dx  \label{rhs}
\end{align}
and this holds for any $f\in L^2(\R^3)$ with $\hat f$ having compact
support. Taking the infimum over all such $f$'s on the right side of
(\ref{rhs}) gives $\Ee(M)$, hence we indeed conclude that
$$
\Ee(\widetilde M) \leq \Ee(M)\,.
$$

\section{Outlook and Open Problems}\label{sec:end}

The results reported here treat only the first of several related questions. Some of these
will occur to the reader and some of them are stated here.

$\bullet$ Conjecture: Look at the ground state of $H +
\lambda^2 V(\lambda x)$ in $\Hh$ for small $\lambda$ and set $x = y/\lambda$ with $y$
fixed. This is a state in Fock space $\F$. When appropriately normalized,  it should converge to $\Phi_0$
(translated by $x$, i.e., conjugated with the unitary $e^{ix P_f}$) as $\lambda\to 0$, irrespective of $y$. 

$\bullet$ What is the relation of $\Phi_0$ to Pekar's variational
minimizer for the translation invariant problem
when $\alpha$ is large? We recall that this vector in $\Hh$ is a
simple tensor product of a vector $\varphi$ in $L^2(\R^3)$ and a
vector (coherent state) $\Phi_{\rm Pekar} \in \F$.  Conjecture: For
large $\alpha$, the ground state $\Phi_0$ of $H_P$ at $P=0$ is close
to
$$
\hat\varphi(P_f) \Phi_{\rm Pekar}
$$
where $P_f$ is the momentum operator on $\F$ and $\hat\varphi(P_f)$ is
the Fourier transform of $\varphi$, as an operator on $\F$. On the
level of $n$-phonon distributions this should be provable using the
analysis in \cite{LT}.

$\bullet$  An increase of $\alpha$ can lead
to a bound state of the system, even in the absence of any bound state
for $\alpha=0$. This \lq\lq quantum phase transition\rq\rq\ can be studied by our
method, by introducing a potential $V$ such that the operator
$p^2/(2m) + V(x)$ has a bound state only for $m > m^* > 1/2$. In order
to study the precise dependence of this transition on $V$ and $\alpha$,
it is necessary to get quantitative bounds on the effective mass,
which is a long-standing open problem.

$\bullet$  How can one describe the corrections to the identity
(\ref{MR}) for small but non-zero $\lambda$? These corrections lead to
physically relevant effects, like the Lamb shift in quantum
electrodynamics.

\bigskip\bigskip

\noindent {\it Acknowledgments.}  Partial financial support by U.S. NSF grant PHY-0965859
(E.H.L.), the Simons Foundation (\# 230207, E.H.L.) and the NSERC (R.S.) is gratefully acknowledged.

\end{document}